\documentstyle[referee,psfig]{mn}
\newif\referee
\newif\ifAMStwofonts

\catcode`\@=11
\def\gsim{\ifmmode{\mathrel{\mathpalette\@versim>}}
    \else{$\mathrel{\mathpalette\@versim>}$}\fi}
\def\lsim{\ifmmode{\mathrel{\mathpalette\@versim<}}
    \else{$\mathrel{\mathpalette\@versim<}$}\fi}
\def\@versim#1#2{\lower 2.9truept \vbox{\baselineskip 0pt \lineskip
    0.5truept \ialign{$\m@th#1\hfil##\hfil$\crcr#2\crcr\sim\crcr}}}
\catcode`\@=12
\def\Au{A_1}
\def\Bu{B_1}
\def\Cu{C_1}
\def\Du{D_1}
\def\Ad{A_2}
\def\Bd{B_2}
\def\Cd{C_2}
\def\Dd{D_2}
\def\Kv{K_{\rm V}}
\def\ml{\Upsilon_*}
\def\Mstar{M_*}
\def\Psistar{\Psi_*}
\def\rhostar{\rho_*}
\def\rhoM{\rho_{\rm M}}
\def\Lb{L_{\rm B}}
\def\Lbsol{L_{\rm B\odot}}
\def\Msol{M_{\odot}}
\def\Ie{\langle I\rangle _{\rm e}}
\def\Re{R_{\rm e}}
\def\cRe{\langle R\rangle _{\rm e}}
\def\Dt{\Delta t}
\def\Dtmax{\Delta t_{\rm max}}
\def\Dtmin{\Delta t_{\rm min}}
\def\sae{a_{\rm e}}
\def\sbe{b_{\rm e}}
\def\rc{r_{\rm c}}
\def\ra{r_{\rm a}}
\def\rme{r_{\rm M}}
\def\sa{s_{\rm a}}
\def\sas{s_{\rm as}}
\def\sac{s_{\rm ac}}
\def\sg0{\sigma_0}
\def\gm{\gamma}
\def\Td{T_{\rm dyn}}
\def\Tr{T_{\rm r}}
\def\Tt{T_{\rm t}}
\def\caf{(c/a)_{\rm fin}}
\def\xis{\xi_{\rm s}}
\def\en{{\mathcal{E}}}
\def\psit{\Psi_{\rm T}}
\def\ku{k_1}
\def\kd{k_2}
\def\kt{k_3}
\def\kti{\kt^{\rm i}}
\def\ktiso{\kt^{\rm iso}}
\def\kuiso{\ku^{\rm iso}}
\def\vr{v_{\rm r}}
\def\vtheta{v_{\vartheta}}
\def\vphi{v_{\varphi}}
\def\nua{\nu_{\rm a}}
\def\alphatol{\alpha_{\rm tol}}
\ifoldfss
  \ifCUPmtlplainloaded \else
    \NewTextAlphabet{textbfit} {cmbxti10} {}
    \NewTextAlphabet{textbfss} {cmssbx10} {}
    \NewMathAlphabet{mathbfit} {cmbxti10} {} 
    \NewMathAlphabet{mathbfss} {cmssbx10} {} 
  \fi
  \ifAMStwofonts
    \ifCUPmtlplainloaded \else
      \NewSymbolFont{upmath} {eurm10}
      \NewSymbolFont{AMSa} {msam10}
      \NewMathSymbol{\upi}     {0}{upmath}{19}
      \NewMathSymbol{\umu}     {0}{upmath}{16}
      \NewMathSymbol{\upaellitrtial}{0}{upmath}{40}
      \NewMathSymbol{\leqslant}{3}{AMSa}{36}
      \NewMathSymbol{\geqslant}{3}{AMSa}{3E}

      \let\leq=\leqslant 
      \let\geq=\geqslant 
    \fi
  \fi
\fi 

\ifnfssone
  \newmathalphabet{\mathit}
  \addtoversion{normal}{\mathit}{cmr}{m}{it}
  \addtoversion{bold}{\mathit}{cmr}{bx}{it}
  \newmathalphabet{\mathbfit} 
  \addtoversion{normal}{\mathbfit}{cmr}{bx}{it}
  \addtoversion{bold}{\mathbfit}{cmr}{bx}{it}
  \newmathalphabet{\mathbfss} 
  \addtoversion{normal}{\mathbfss}{cmss}{bx}{n}
  \addtoversion{bold}{\mathbfss}{cmss}{bx}{n}
  \ifAMStwofonts
    \ifCUPmtlplainloaded \else
      \UseAMStwoboldmath
      \makeatletter
      \new@mathgroup\upmath@group
      \define@mathgroup\mv@normal\upmath@group{eur}{m}{n}
      \define@mathgroup\mv@bold\upmath@group{eur}{b}{n}
      \edef\UPM{\hexnumber\upmath@group}
      \new@mathgroup\amsa@group
      \define@mathgroup\mv@normal\amsa@group{msa}{m}{n}
      \define@mathgroup\mv@bold\amsa@group{msa}{m}{n}
      \edef\AMSa{\hexnumber\amsa@group}
      \makeatother
      \mathchardef\upi="0\UPM19
      \mathchardef\umu="0\UPM16
      \mathchardef\upartial="0\UPM40
      \mathchardef\leqslant="3\AMSa36
      \mathchardef\geqslant="3\AMSa3E

      \let\leq=\leqslant 
      \let\geq=\geqslant 
    \fi
  \fi
\fi 

\ifnfsstwo
  \DeclareMathAlphabet{\mathbfit}{OT1}{cmr}{bx}{it}
  \SetMathAlphabet\mathbfit{bold}{OT1}{cmr}{bx}{it}
  \DeclareMathAlphabet{\mathbfss}{OT1}{cmss}{bx}{n}
  \SetMathAlphabet\mathbfss{bold}{OT1}{cmss}{bx}{n}
  \ifAMStwofonts
    \ifCUPmtlplainloaded \else
      \DeclareSymbolFont{UPM}{U}{eur}{m}{n}
      \SetSymbolFont{UPM}{bold}{U}{eur}{b}{n}
      \DeclareSymbolFont{AMSa}{U}{msa}{m}{n}
      \DeclareMathSymbol{\upi}{0}{UPM}{"19}
      \DeclareMathSymbol{\umu}{0}{UPM}{"16}
      \DeclareMathSymbol{\upartial}{0}{UPM}{"40}
      \DeclareMathSymbol{\leqslant}{3}{AMSa}{"36}
      \DeclareMathSymbol{\geqslant}{3}{AMSa}{"3E}

      \let\leq=\leqslant 
      \let\geq=\geqslant 
    \fi
  \fi
\fi 

\ifCUPmtlplainloaded \else
  \ifAMStwofonts \else 
    \def\upi{\pi}
    \def\umu{\mu}
    \def\upartial{\partial}
  \fi
\fi

\title{Radial orbital anisotropy and the Fundamental
       Plane of elliptical galaxies}
\author[C. Nipoti, P. Londrillo and L. Ciotti]
  {C.~Nipoti,$^1$ P.~Londrillo$^2$ and L.~Ciotti$^{2,3,4}$\\
  $^1$Dipartimento di Astronomia dell'Universit\`a degli Studi di Bologna, 
      via Ranzani 1, 40127 Bologna, Italy\\
  $^2$Osservatorio Astronomico di Bologna, 
      via Ranzani 1, 40127 Bologna, Italy\\
  $^3$Scuola Normale Superiore, 
      Piazza dei Cavalieri 7, 56126 Pisa, Italy\\
 $^4$Department of Astrophysical Sciences, 
      Princeton University Observatory, Ivy Lane, NJ 08544 USA}
  
\date{Revised version, December 17, 2001}

\pagerange{\pageref{firstpage}--\pageref{lastpage}}
\pubyear{2001}

\begin{document}

\maketitle

\label{firstpage}

\begin{abstract}

The existence of the Fundamental Plane imposes strong constraints on
the structure and dynamics of elliptical galaxies, and thus contains
important information on the processes of their formation and
evolution.  Here we focus on the relations between the Fundamental
Plane thinness and tilt and the amount of radial orbital anisotropy:
in fact, the problem of the compatibility between the observed
thinness of the Fundamental Plane and the wide spread of orbital
anisotropy admitted by galaxy models has been often risen. By using
N-body simulations of galaxy models characterized by observationally
motivated density profiles, and also allowing for the presence of
live, massive dark matter halos, we explore the impact of radial
orbital anisotropy and instability on the Fundamental Plane
properties.  The numerical results confirm a previous semi--analytical
finding (based on a different class of one--component galaxy models):
the requirement of stability matches almost exactly the thinness of
the Fundamental Plane.  In other words, galaxy models that are
radially anisotropic enough to be found outside the observed
Fundamental Plane (with their isotropic parent models lying on the
Fundamental Plane) are unstable, and their end--products fall back on
the Fundamental Plane itself.  We also find that a systematic increase
of radial orbit anisotropy with galaxy luminosity cannot explain by
itself the whole tilt of the Fundamental Plane, becoming the galaxy
models unstable at moderately high luminosities: at variance with the
previous case their end--products are found well outside the
Fundamental Plane itself. Some physical implications of these findings
are discussed in detail.

\end{abstract}


\section{Introduction}

The Fundamental Plane (FP) of elliptical galaxies (Djorgovsky \&
Davies 1987; Dressler et al. 1987) is a scaling relation between three
of their basic {\it observational} properties, namely the circularized
effective radius $\cRe\equiv\sqrt{\sae\sbe}$ (where $\sae$ and
$\sbe$ are the major and minor semi-axis of the effective isophotal
ellipse), the central velocity dispersion $\sg0$, and the mean
effective surface brightness $\Ie{\equiv}\Lb/2\pi{\cRe}^2$ (where
$\Lb$ is the luminosity of the galaxy, for example in the Johnson
B-band).  An interesting parameterization of the FP, that we adopt
in this paper, has been introduced by Bender, Burstein \& Faber (1992,
hereafter BBF):
\begin{equation}
\ku\equiv\frac{\log\sg0^2 + \log\cRe}{\sqrt 2},
\end{equation}
\begin{equation}
\kd\equiv\frac{\log\sg0^2 + 2\log\Ie - \log\cRe}{\sqrt 6},
\end{equation}
\begin{equation}
\kt\equiv\frac{\log\sg0^2 - \log\Ie -\log\cRe}{\sqrt 3}.
\end{equation}
In particular, when projected on the $(\ku ,\kt)$ plane, the FP is
seen almost edge--on and it is considerably thin, while the
distribution of galaxies in the $(\ku ,\kd)$ plane is considerably
broader.  For example, Virgo ellipticals studied by BBF are
distributed on the $(\ku ,\kt)$ plane according to the best--fit
relation
\begin{equation}
\kt=0.15\ku+0.36
\end{equation}
(when adopting respectively, kpc, km s$^{-1}$ and $\Lbsol$ pc$^{-2}$
as length, velocity and surface brightness units), with
a very small dispersion of $\sigma (\kt)\simeq 0.05$ over all the
range spanned by the data, $2.6\, \lsim\, \ku\, \lsim\, 4.6$ (and so
$0.75\, \lsim \, \kt\, \lsim \, 1.05$, see, e.g., Ciotti, Lanzoni \&
Renzini 1996, hereafter CLR96).

By combining equations (1) and (3) with equation (4) the FP equation
of BBF is then obtained directly in terms of the observables: the
exponents are in good agreement with those derived (in the
Johnson B-band) from a much larger galaxy sample by J{\o}rgensen,
Franx \& Kj{\ae}rgaard (1996).

Note that equation (4) implies that for galaxies of given luminosity
$\Lb$ their effective radius and central velocity dispersion must be
strongly coupled: in fact at any  fixed
luminosity the coordinates $\kt$ and $\ku$ are related through
definitions (1) and (3) by
\begin{equation}
\kt=\sqrt{\frac{2}{3}}\ku+{\sqrt{\frac{1}{3}}}\log {\frac{2\pi}{\Lb}},
\end{equation}
and the slope of this relation ($\simeq 0.82$) is different from that
of the FP ($\simeq 0.15$). As a consequence, in the $(\ku,\kt)$ plane
all galaxies with the same luminosity are located on straight lines
significantly inclined with respect to the FP: the presence of
substantial scatter in galactic properties from galaxy to galaxy (of
similar luminosity) would destroy the thinness of the FP by producing
a large scatter in $\ku$ and so in $\kt$.

The relation between galaxy properties and the FP can be expressed in
a quantitative way under the reasonable assumption that present day
ellipticals are virialized systems. We write the virial theorem as
\begin{equation}
\frac{G\Lb\ml}{\cRe}=\Kv\sg0^2 
\end{equation}
where $\ml=\Mstar/\Lb$ is the galaxy {\it stellar} mass--to--light
ratio (for example in Blue solar units), and $\Kv$ is a dimensionless
factor depending on the stellar density profile, internal dynamics,
dark matter amount and distribution, and, for non--spherical galaxies,
on their relative orientation with respect to the observer's
line--of--sight (see, e.g., Ciotti 1997); in addition, $\Kv$ depends
also on the observing aperture adopted to derive $\sg0$.

Equations (4) and (6) imply that, for galaxies belonging to the FP,
the quantity $\ml/\Kv$ is a very well defined function of
two\footnote{In principle one could find virialized galaxies {\it
everywhere} in the three--dimensional observational ($\Lb, \sg0,
\cRe$) space.  From this point of view the existence of the FP is
related to the virial theorem as the HR diagram is related to
hydrostatic equilibrium: the useful information that one derives is
{\it not} about the equilibrium equations, but the physics of the
objects involved.} of the three observables $\Lb$, $\sg0$, and
$\cRe$. In the particular case of adopting the numerical coefficients
of equation (4), measuring $\Lb$ in
$10^{10}\Lbsol$, and taking into account that $\ku=\log
(G\ml\Lb/\Kv)/\sqrt{2}$ and $\kt=\log (2\pi G\ml/\Kv)/\sqrt{3}$,
\begin{equation}
\frac{\ml}{\Kv}\simeq 1.12\times{\Lb}^{0.23}
\end{equation}
where the quantity $\ml/\Kv$ is characterized by a scatter of $\simeq
20\%$ due to its relation with $\kt$ (however, additional
considerations reduce this figure to $\simeq 12\%$, see Renzini \&
Ciotti 1993).  As a consequence, any departure from the relation
dictated by equation (7) will move a galaxy away from the FP.  We
recall here that the dependence of $\kt$ on $\ku$, as given by
equation (4) and responsible for the luminosity dependence of the
ratio $\ml/\Kv$, is commonly known as the FP ``tilt''.

The simple analysis presented above shows that the two properties of
thinness and tilt of the FP are deeply connected with the present--day
structure and dynamics of ellipticals (hereafter, Es), and, as a
consequence, with their formation and evolution history: the very
existence of the FP [as well as of the other tight scaling relations
revealing the remarkable homogeneity of Es, such as the $M_{\rm
BH}-\sg0$ (Gebhardt et al.  2000; Ferrarese \& Merritt 2000), the
${\rm Mg}_2$-$\sg0$ (Bender, Burstein \& Faber, 1993 and references
therein) and the color--magnitude (Bower, Lucey \& Ellis, 1992)
relations] imposes strong constraints on the different formation and
evolutionary scenarios proposed for Es (i.e., dissipationless merging,
monolithic collapse, or a combination of the two; see, e.g., Ciotti \&
van Albada 2001).

Among the various galaxy properties in principle able to destroy the
FP thinness (as a consequence of a substantial variation at fixed
galaxy luminosity), one of the most ``effective'' is certainly orbital
anisotropy (de Zeeuw \& Franx 1991). In fact, galaxy models are
commonly believed to be able to sustain a large spread of orbital
anisotropies and it is also well known that radial orbital anisotropy
can produce very high {\it central} velocity dispersion values, and
correspondingly low values of $\Kv$, thus substantially violating
equation (7).  A natural question to be addressed is then what
physical principle or evolutionary process limits the range of orbital
anisotropies shown by real galaxies.  Ciotti \& Lanzoni (1997,
hereafter CL97), using one--component, radially anisotropic Sersic
(1968) models, and a semi--analytical investigation based on the
Fridman \& Polyachenko (1984, hereafter FP84) stability indicator,
suggested the possibility that radial orbit instability could be the
limiting factor of the FP thickness.  In practice, CL97 found
indications that galaxy models sufficiently anisotropic to be outside
the FP observed thickness (when their parent isotropic model was
assumed to lie on the FP) were unstable.  Clearly, this preliminary
indication requires a confirmation with the aid of numerical
simulations and more realistic galaxy models (for example, allowing
for the presence of live dark matter halos). Also, a question
naturally associated with that above is the determination of the
position of the end--products of unstable initial conditions in the
space of the observables.  One of the aims of this work is indeed to
answer these two questions by numerical simulations of one and
two--component radially anisotropic galaxy models.

Orbital anisotropy is not only related to the problem of the FP
thinness but it is also one of the candidates that have been proposed
to explain the origin of the FP tilt (CLR96; CL97). If this were the
case, the amount of radial anisotropy in the velocity dispersion
tensor should increase with galaxy luminosity, as can be seen from
equation (7) under the assumption of a constant $\ml$ and of
structural homology\footnote{With ``structural homology'' we mean that
all the structural galaxy properties (e.g., the stellar and dark halo
density profiles, the ratio of their scale--lengths and masses, and so
on) {\it do not depend} on $\Lb$.} over the whole FP plane.  In other
words, in this scenario the FP tilt would be produced by a {\it
dynamical} non--homology due to anisotropy (note that dynamical
non--homology may well coexist with structural homology, but the
converse is in general not true). Many interesting questions are
raised by the scenario depicted above: for example, under the
assumption that an isotropic galaxy of given luminosity lies on the
FP, how far can the derived structurally homologous but radially
anisotropic models climb over the FP before the onset of radial orbit
instability?  In addition, what happens to the end--products of the
unstable models?  Will they remain near the FP?  In this paper we try
to address also these questions with the aid of N-body numerical
simulations.

Strictly related to the clarification of the interplay between orbital
anisotropy and the FP tilt, is the possibility to obtain some clues on
the formation processes of Es.  In fact it is trivial to prove that,
as consequence of the virial theorem and the conservation of the total
energy, in the merging of two galaxies with masses $M_1$ and $M_2$ and
virial velocity dispersions $\sigma_{\rm v,1}$ and $\sigma_{\rm v,2}$, the
virial velocity dispersion of the resulting galaxy is given by
\begin{equation}
\sigma^2_{\rm v,1+2}={M_1\sigma^2_{\rm v,1}+M_2\sigma^2_{\rm v,2}\over M_1+M_2}
\end{equation}
(by definition $\sigma^2_{\rm v}\equiv 2T/M$, where $T$ is the total
kinetic energy of the galaxy).  For simplicity in the formula above we
considered, in the initial conditions, a negligible energy of the
galaxy pair when compared to the other energies involved in the
process, and no significant mass loss from the resulting system. From
equation (8) it follows that $\sigma_{\rm v,1+2}\leq\,{\rm
max}(\sigma_{\rm v,1},\sigma_{\rm v,2})$, i.e., the {\it virial}
velocity dispersion cannot increase in a merging process of the kind
described above. On the other hand, the FP (or the less tight
Faber--Jackson relation; Faber \& Jackson 1976) indicates that the
{\it projected, central} velocity dispersion increases with galaxy
luminosity.  Then, in the dissipationless merging scenario the FP tilt
can be produced only by structural and/or dynamical non--homology,
since the relation between central and virial velocity dispersion
depends on the structure and dynamics of the system: in particular we
will discuss here the second possibility, with regard to an increase
in radial orbital anisotropy with galaxy luminosity.  Note that an
increase in the radial orbit amount has been claimed in the past as a
natural by--product of galaxy merging, and also some observations have
been interpreted in this way (Bender 1988; see also Naab, Burkert \&
Hernquist 1999, and references therein). With the aid of the explored
numerical models we will try to obtain some qualitative insight in
this problem: it is however clear that the results should be
considered at the best qualitative indications, and that a firm answer
about the role of merging in producing the FP tilt can be obtained
only with N-body numerical simulations of merging galaxies (see, e.g.,
Capelato, de Carvalho \& Carlberg 1995).

Summarizing, the aims of this work are the following. For what
concerns the FP thickness problem, we investigate the role of radial
orbit instability as a factor regulating the amount of radial
anisotropy for galaxies of given luminosity, and the position,
relative to the FP, of the end--products of radially unstable
anisotropic models. In addition, we determine whether it is possible
to reproduce the whole FP tilt with a systematic variation of radial
anisotropy with luminosity (using both stable and unstable initial
conditions), and what is the fate of unstable models initially forced
to lie on the FP.  This paper is organized as follows.  In Section 2
we describe the basic structural and dynamical properties of the
investigated models; a short description of the codes used for the
numerical simulations is also given.  In Section 3 we present the
results and their impact on the FP thinness problem, and in Section 4,
the results are discussed focusing on the origin of the FP tilt.
Finally, in Section 5 the main conclusions are summarized.

\section{Models}

\subsection{Initial conditions}

\subsubsection{Structural and dynamical properties}

As initial conditions for the N-body simulations we use spherically
symmetric one--component $\gm$-models (Dehnen 1993; Tremaine et
al. 1994) and two--component ($\gm_1,\gm_2$) models (Ciotti 1996,
1999).  The density, mass, and (relative) potential
profiles of the stellar component are given by
\begin{equation}
\rhostar (r)= {3-\gm\over 4\pi}{\Mstar\rc \over r^{\gm} (\rc+r)^{4-\gm}}  
\qquad  (0\leq \gm <3),
\end{equation}
\begin{equation}
{\Mstar (r)\over\Mstar}=\left({r\over \rc+r }\right)^{3-\gm},
\end{equation}
\begin{equation}
{\Psistar (r)}={G\Mstar \over \rc (2-\gm)}
            {\left[1-\left({r\over \rc+r}\right)^{2-\gm}\right]} 
            \quad (\gm\,\neq\,2),
\end{equation}
\begin{equation}
{\Psistar (r)}={G\Mstar \over \rc}{\ln {\rc+r\over r}} \qquad (\gm=2),
\end{equation}
where $\Mstar$ is the total stellar mass; the $\gm=1$ and $\gm=2$
cases correspond to Hernquist (1990) and Jaffe (1983) density
distributions, two reasonable approximations (when projected) of the
$R^{1/4}$ law (de Vaucouleur 1948).  In the two--component models, the
dark matter (DM) halo is described by $\rho_{\rm h}$,
$M_{\rm h}$ and $\Psi_{\rm h}$ profiles of the same family of
equations (9)-(12), where now $r_{\rm h}\equiv{\beta}\rc$ and $M_{\rm
h}\equiv{\mu}\Mstar$.

Radial anisotropy in the stellar orbital distribution is introduced by
using the Osipkov-Merritt (OM) parameterization (Osipkov 1979; Merritt
1985).  The distribution function (DF) of the stellar component is
then given by
\begin{equation}
f_*(Q)=\frac{1}{\sqrt{8}\pi^2}\frac{d}{dQ}
       \int_0^Q{\frac{d{\varrho_*}}{d\psit}}{\frac{d\psit}{\sqrt{Q-\psit}}},
\end{equation}
where
\begin{equation}
\varrho_* (r)=\left(1+\frac{r^2}{\ra^2}\right)\rhostar (r).
\end{equation}
The variable $Q$ is defined as $Q\equiv \en-{L^2/2\ra^2}$, where the
relative (positive) energy is given by $\en =\psit-v^2/2$, $v$ is the
modulus of the velocity vector, the relative total potential is
$\psit=\Psistar +\Psi_{\rm h}$, $L$ is the angular momentum modulus
per unit mass, and $f_*(Q)=0$ for $Q\leq0$. The quantity $\ra$ is the
so--called ``anisotropy radius'': for $r \gg\ra$ the velocity
dispersion tensor is mainly radially anisotropic, while for $r \ll \ra
$ the tensor is nearly isotropic. Isotropy is realized at the model
center, independently of the value of $\ra$; in the limit $\ra
\to\infty$, $Q=\en$ and the velocity dispersion tensor becomes
globally isotropic.

In order to reduce the dimensionality of the parameter space,
the orbital distribution of the DM halos is assumed isotropic in all
our simulations; as a consequence the DF for the DM
halo component, $f_{\rm h}(\en)$, is given by equation (13) where
$Q=\en$ and $\rho_{\rm h}(r)$ substitutes $\varrho_*(r)$.
  
\subsubsection{Physical scales}

According to the given definitions, from the structural and
dynamical point of view the one--component models are completely
determined by four quantities: the two physical
scales $\Mstar$ and $\rc$, and the two dimensionless parameters $\gm$
and $\sa \equiv\ra/\rc$.

In the numerical simulations $\Mstar$, $\rc$ and $\Td$ are adopted as
mass, length and time scales.  $\Td$ is the half--mass dynamical time,
defined as
\begin{equation}
\Td \equiv \sqrt{\frac{3\pi}{16G\rhoM}} =
           \pi\sqrt{\frac{\rc^3}{2G\Mstar}}F(\gm),
\end{equation}
where $\rhoM =3\Mstar /8\pi \rme^3$ is the mean density inside the
half--mass radius $\rme$ and
\begin{equation}
F(\gm)=\left(2^{\frac{1}{3-\gm}}-1\right)^{-\frac{3}{2}}.
\end{equation}
Note that at fixed $\Mstar$ and $\rc$ the half--mass dynamical time
depends strongly on $\gm$.  For example, $F(0)/F(1)\simeq 2.0$ and
$F(1)/F(2)\simeq 3.8$. Finally, the velocity scale is given by
\begin{equation}
v_{\rm c}\equiv\frac{\rc}{\Td}=\frac{1}{\pi F(\gm)}
                               \sqrt{\frac{2G\Mstar}{\rc}}.
\end{equation}

In the case of two--component galaxies we limit our present study to
(1,1) models, i.e., to two--component Hernquist models (Ciotti
1996). In this way the DM halo is similar to Navarro, Frenk \& White
(1996) profiles.  (1,1) models are characterized by five quantities,
the two physical scales $\Mstar$ and $\rc$, and the three
dimensionless parameters $\sa$, $\mu$ and $\beta$. Their $\Td$
depends on $\mu$ and $\beta$: from definition (15), where now $\rhoM =
3(1+\mu)\Mstar/(8\pi\rme^3)$ and $\rme$ is the half--mass radius of the
total (stellar plus dark) density distribution, it results that 
\begin{equation}
\Td=\pi\sqrt{\frac{\rc^3}{2G\Mstar}}F(\mu,\beta),
\end{equation}
where $F=x^{3/2}(\mu,\beta)/\sqrt{1+\mu}$, and $x$ is the
 solution of
\begin{equation}
\left({x \over 1+x}\right)^2+\mu\left({x\over\beta+x}\right)^2
={1+\mu\over 2}.
\end{equation}

Finally note that, once the dimensionless numbers $\sa$ and $\gm$ (for
one--component models) and $\sa$, $\mu$, and $\beta$ (for
two--component models) are fixed, the results of the numerical
simulations can be rescaled for arbitrary values of $\Mstar$, $\rc$,
and $\Lb$ (or $\ml$).  In particular, in order to transform the results
of the numerical simulations in physical units we express $\Mstar$ in
$10^{10}\Msol$, $\rc$ in kpc and $\Lb$ in $10^{10}\Lbsol$.
 
\subsubsection{Numerical realization of the initial conditions}

In order to arrange the initial conditions for the numerical
simulations, we distribute $N$ particles ($N=32768$ and $N=131072$ for
one and two--component models, respectively) by using spherical
coordinates ($r,\vartheta,\varphi$, for positions, and
$\vr,\vtheta,\vphi$, for velocities). The radial coordinate is assigned
by inverting equation (10) and choosing a sampling suited to resolve
the core of the distribution, while angular coordinates are given
randomly. The velocity vector of each particle is assigned by using
the von Neumann rejection method. This method can be easily
applied to OM systems by introducing the dimensionless vector of
components $(u_1 ,u_2 ,u_3)$, related to $(\vr, \vtheta ,\vphi)$ by
\begin{equation}
\vr     = \sqrt{2\psit}u_1, \quad
\vtheta = \frac{\sqrt{2\psit}}{\nua}u_2, \quad
\vphi   = \frac{\sqrt{2\psit}}{\nua}u_3, 
\end{equation}
where
\begin{equation}
\nua =\sqrt{1+\frac{r^2}{\ra^2}};  \\
\end{equation}
from this choice $0\leq u\leq1$, where $u=\sqrt{u_1^2+u_2^2+u_3^2}$.
The vector $(u_1, u_2, u_3)$ is assigned to each particle by using the
rejection method, after computing the value of the DF at the
required $Q=(1-u^2)\psit$, by numerically evaluating equation (13).
The components of the velocity vector $(\vr ,\vtheta, \vphi)$ are then
recovered according to equation (20).  The initial conditions so
obtained are then compared to the expected analytical density
distribution, finding, as a rule, an agreement of better than 2\% over
the radial interval containing 0.99 of the total mass of the
theoretical model.

\subsection{The numerical codes}

For our simulations we used the Barnes \& Hut (1986) TREECODE (in a
version made publicly available by Hernquist, 1987) and the Springel,
Yoshida \& White (2000) parallel version of GADGET (adapted to run on
the 32 processors Cray T3E at CINECA).  In particular,  the
one--component simulations were run on a Alpha workstation by using the
TREECODE with $N=32768$; for some of them we used also $N=65536$.  The
quadrupole correction in the cell--particle force calculation was
always applied.  For the two--component models we used instead GADGET
with $N=131072$; in addition, some one--component models were also
tested with $N=131072$ using the same code.  We found very good
agreement between the test simulations performed with both the codes,
with the basic properties of the simulated systems nearly independent
of the adopted number of particles.

From the numerical point of view, TREECODE simulations are
characterized by three parameters, namely the opening angle $\theta$,
the softening parameter $\varepsilon$ (i.e., the softening length
expressed in units of $\rc$) and the time step $\Dt$.  Following
Hernquist (1987) and Barnes \& Hut (1989), we assume $\theta=0.8$, a
good compromise between conservation of total energy and computational
time.  The choice of $\Dt$ and $\varepsilon$ requires some care; in
fact these two parameters are strongly coupled and, in order to
maintain the same accuracy in the force evaluation, $\Dt$ must be
reduced if $\varepsilon$ is reduced (Barnes \& Hut 1989).  In
addition, the ``optimal'' value of the softening length is strongly
dependent on $N$ and on the specific density distribution profile
(Merritt 1996; Athanassoula et al. 2000; Dehnen 2001).  To choose the
values of these parameters, we performed some simulations of isotropic
$\gm$-models over 100 $\Td$, checking the total energy $E$ and virial
ratio ($V=2T/W$) conservation.  On the basis of these tests we adopt
$\Dt =\Td/100$, and $\varepsilon =(0.072,0.030,0.016)$ for
$\gm=(0,1,2)$ models, respectively. With this choice we obtain
$|\Delta E /E|<1.5\%$ and $|\Delta V/V|<2\%$.  GADGET simulations
depend on five parameters: the cell--opening parameter $\alpha$, the
minimum and the maximum time step $\Dtmin$ and $\Dtmax$, the
time--step tolerance parameter $\alphatol$, and the softening
parameter $\varepsilon$ (Springel et al. 2000).  Fiducial values of
the parameters were fixed after running a few one--component ($\gm=1$)
test simulations.  In particular, with $\alpha=0.02$, $\Dtmin =0$,
$\Dtmax =\Td/100$, $\alphatol =0.05$, and $\varepsilon =0.016$ we
obtain $|\Delta E/E |<0.5\%$ and $|\Delta V/V|<1\%$.  The value of
the first four parameters were used also in the two--component
simulations. The softening length $\varepsilon$ was instead determined
case by case, by considering the core radius length of the more
concentrated component.

In order to determine the ``observational'' properties of the
end--products of the numerical simulations, and place them in the
$(\ku ,\kd ,\kt)$ space, we measured their effective radius $\cRe$,
central velocity dispersion $\sg0$ and mean effective surface
brightness $\Ie$ for several projection angles.  The technical details
of the adopted procedures are given in the Appendix, together with a
brief discussion on the discreteness effects on the derived values of
the ``observational'' properties of the models. A point of interest is
the choice of the adopted ``aperture radius'' for the measure of
$\sg0$, fixed in this work to $\cRe /8$ in order to match the
observational procedure at low redshifts (see, e.g., J{\o}rgensen et
al. 1996). Indeed, it is well known that isotropic $\gamma$-models may
present a projected velocity dispersion profile decreasing toward the
center (in contrast to what happens in the vast majority of elliptical
galaxies): for example, the projected velocity dispersion of Hernquist
models peaks approximately at $\Re/5$, while the profile for
(isotropic) Jaffe models is monotonically decreasing.  As a
consequence, the position of the initial conditions and of the
end--products in the $k$ space could depend on the adopted aperture
radius used to measure $\sg0$. In order to assess the effect of this
choice on our conclusions, we analyzed the results of the simulations
also by using an aperture radius of $\cRe /4$, and we found very good
agreement with the results obtained with the aperture $\cRe/8$:
therefore, we present here only these last results. Note however that,
by virtue of the projected virial theorem (see, e.g., Ciotti 1994), a
totally different scenario from that explored in this paper would
arise in the limiting case of $\sg0$ measured over the whole galaxy:
anisotropy would play no role at all in determining the position of
initial conditions in the $k$ space, and any difference between
initial and final $\sg0$ would be only due to projection effects
associated with loss of spherical symmetry. These aperture effects
could be important when studing observationally the FP at intermediate
redshift (see, e.g., van Dokkum \& Franx 1996, Bender et al. 1998,
Pahre, Djorgovski \& de Carvalho 1998, Treu et al. 1999), or locally
by using large apertures (see, e.g., Graham \& Colless 1997).

\section{Orbital anisotropy and the FP thinness}

As already pointed out in the Introduction, the 1--sigma dispersion of
the observational data around the best--fit relation (4) is
surprisingly small and nearly constant over all the observed range in
$\ku$, with $\sigma (\kt)\simeq 0.05$.  We investigate here the
constraints imposed by this tightness on the amount of radial orbital
anisotropy for the set of one and two--component galaxy models
described in Section 2.  We start by fixing the values for the
dimensionless parameters $\gm$ (for one--component models) and $\mu$
and $\beta$ (for two--component models); we also assume global
isotropy, i.e., $\sa\to\infty$, and in this way the quantity $\Kv$ is
uniquely determined. These globally isotropic models (that we call
{\it parent models}) are then placed on the FP by assigning of
the pair $(\ml,\Lb)$ so that equation (7) is verified.  From each of
these parent models lying on the FP we then generate a {\it family} of
OM radially anisotropic models by decreasing $\sa$, while maintaining
fixed all the other model parameters.  Correspondingly $\Kv$ decreases
(as can be seen from Table 1, in the case of one--component models and
for representative values of the parameter $\sa$), $\ku$ increases,
and so does $\kt$, according to equation (5): for sufficiently small
values of $\sa$ the members of each family are found outside the observed
thickness of the FP. Note that by an appropriate choice of the pair
$(\ml ,\Lb)$ each parent isotropic model can be placed at arbitrary
positions over the best--fit line (4), and so the results of the
numerical simulations (after a rescaling to $\ml$ and $\Lb$) are the
same everywhere on the FP.  In addition, since $\sigma (\kt)$ is
constant over the whole observational range spanned by $\ku$, the
conclusions obtained from each family of models are also independent
of the position of the parent galaxy on the FP.
\begin{table}
 \centering
  \caption
{The dimensionless coefficient $\Kv(\gm,\sa)$ for one--component
models, as obtained from equation (6). See Section 3 for the
definitions of $\sac$ and $\sas$.}
  \begin{tabular}{@{}cccc@{}}
    $\gm$   &   $\Kv(\gm,\infty)$    & $\Kv(\gm,\sas)$ & $\Kv(\gm,\sac)$ \\
    [10pt] 
    0   & 6.7  & 5.9   & 2.8 \\
    1   & 5.9  & 5.3   & 2.7 \\
    2   & 4.8  & 4.4   & 2.3 \\
\end{tabular}
\end{table}

Obviously all isotropic models discussed in this paper are stable
(see, e.g., Binney \& Tremaine 1987; Ciotti 1996), while for each
family of models a critical value $\sas$ for {\it stability} exists
such that the initial conditions characterized by $\sa<\sas$ describe
radially unstable configurations.  From the point of view of the
present discussion, the critical value $\sas$ corresponds, through
$\Kv(\sas)$, to the maximum distance that a {\it stable} model can
have from the FP, where its parent isotropic model lies at (say)
$(\kuiso, \ktiso)$.  Clearly, initial conditions describing unstable
models can be placed at larger distances from the FP: these initial
conditions will evolve with time, and their representative points in
the space of observables will also evolve with time, up to
virialization. In particular, the coordinates $\ku (t)$ and $\kt (t)$
will evolve with time moving on the line described by equation (5).
Note that the maximum distance from the FP at which unstable models
can be placed is in general finite: in fact, the anisotropy radius of
all physically acceptable (stable and unstable) galaxy models must
satisfy the inequality $\sa\geq\sac$, where $\sac\leq\sas$ is the
(dimensionless) critical anisotropy radius for {\it consistency}
(i.e., the anisotropy limit for initial states with a nowhere negative
DF); for a study of $\sac$ in $\gm$--models and in ($\gm_1$,$\gm_2$)
models see Carollo, de Zeeuw \& van der Marel (1995) and Ciotti (1996,
1999).

FP84 argued that a quantitative indication on the maximum amount of
radial orbits sustainable by a specific density profile is given by
the stability parameter $\xi=2 \Tr/\Tt$, where $\Tr$ and $\Tt \equiv
T_{\vartheta}+T_{\varphi}$ are the radial and tangential component of
the kinetic energy tensor, respectively.  From its definition $\xi\to
1$ for $\sa\to\infty$ (globally isotropic models), while
$\xi\to\infty$ for $\sa\to 0$ (fully radially anisotropic models). The
fiducial value indicated by FP84 as a boundary between stable and
radially unstable systems is $\xis\simeq 1.7$. Unfortunately, the
reliability of such indicator is not well understood, and indications
exist of a significant dependence on the particular density profile
under scrutiny (see, e.g., Merritt \& Aguilar 1985; Bertin \&
Stiavelli 1989; Saha 1991, 1992; Bertin et al. 1994; Meza \& Zamorano
1997). In any case, CL97 used this value to determine $\sas$ for
one--component Sersic (1968) models by solving the associated Jeans
equations, and from this value they determined the maximum distance of
stable models from the FP: all models characterized by $\xi < \xis$
where found inside the observed thickness of the FP.  This finding can
be considered at the best a qualitative indication, considering the
uncertainties associated to the exact value of $\xis$ and to its
dependence on the specific density profile adopted, and the need of
numerical simulations is clear.

With the aid of N--body simulations in the following two Sections we
investigate how distant from the FP stable models of various families
can be placed, by increasing their radial orbital anisotropy.  The
logically related question of what happens to the end--products of the
unstable (but physically consistent) initial conditions is also
addressed.

Due to its central role in the following discussion, it is important
to quantify the concept of ``distance'' of a galaxy model from the FP.
In general, we define distance of a point $(\ku ,\kt)$ from the FP the
quantity $\delta\kt\equiv |\kt -0.15\ku -0.36|$, i.e., the distance at
fixed $\ku$ from the point and the FP itself: this quantity naturally
compares with $\sigma (\kt )$.  Unfortunately, as already discussed in
the Introduction, galaxy models of fixed luminosity move along
inclined lines in the $(\ku ,\kt)$ space and so in the present exploration
$|\kt-\ktiso|$ is {\it not} the distance from the FP.  The relation
between $\delta\kt$ and $|\kt-\ktiso|$ is however of immediate
determination: in fact, from equation (5) $\kt =\ktiso+\sqrt{2/3}(\ku
-\kuiso)$, and from the assumption that the parent isotropic models
are placed on the FP one obtains
$\delta\kt=(1-0.15\sqrt{3/2})|\kt-\ktiso|\simeq 0.816|\kt-\ktiso|$.

\subsection{One--component models}

\begin{figure}
\begin{center}
\parbox{1cm}{
\psfig{file=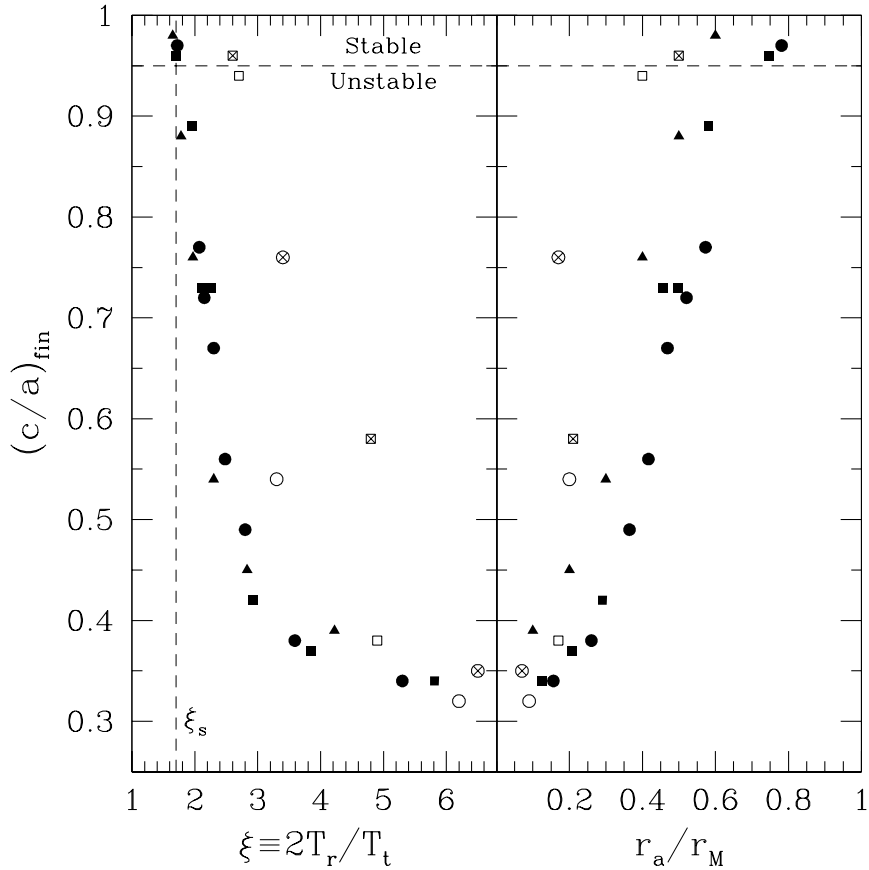,height=12cm}}
\caption{{\it Left}: final axis ratio $\caf$ vs. the stability
parameter $\xi \equiv 2\Tr/\Tt$ of initial conditions, for all the
computed models.  One--component models are represented by full
symbols ($\gm=0$, circles; $\gm=1$, squares; $\gm=2$,
triangles). Two--component (1,1) models are represented by empty
symbols: squares correspond to concentrated halos $(\beta=0.5)$,
circles to diffuse halos $(\beta=2)$. Crosses indicate models with
massive halos $(\mu=3)$. {\it Right}: final axis ratio $\caf$
vs. $\ra/\rme$ for the same models shown in left panel.  The vertical
dashed line is located at $\xis =1.7$ (FP84). Models above the
horizontal dashed line are stable (see the text for a discussion).}
\end{center}
\end{figure}

In order to answer to the questions outlined above, we performed a set
of 21 simulations of one--component $\gm$-models with $\gm=(0,1,2)$,
evaluating numerically for each of the three families the critical
value for stability $\sas$. We recall here that in our simulations the
onset of instability is just due to numerical noise produced by
discreteness effects in the initial conditions, and that, at most, the
simulations are interrupted after 100 $\Td$. As a rule, in order to
determine if a given model is unstable, we found useful to check its
departures from spherical symmetry by monitoring the evolution of its
intrinsic axis ratios $c/a$ and $b/a$ (where $a$, $b$ and $c$ are the
longest, intermediate and shortest axis of its inertia ellipsoid
associated to ``bona fide'' bound particles, see Appendix): we found
that numerical uncertainties (due to the finite number of particles)
of these ratios never exceed $5\%$. According to this choice we define
{\it unstable} the models for which the minimum $c/a$ over 100 $\Td$
is smaller than a fiducial threshold value, 0.95 (the horizontal
dashed line in Fig. 1). Also this value has been obtained by analyzing
the fluctuations -- due to the finite number of particles -- of $c/a$
shown by the numerical realizations of the isotropic (stable) parent
$\gm$-models.  As expected, we found that an exact determination of
$\sas$ for a given density profile is not straightforward: in fact,
while for strongly anisotropic initial conditions the onset of
instability is apparent and the numerical models settle down into a
final equilibrium configuration in a few dynamical times, for nearly
stable initial conditions the instability can be characterized by very
slow growth rates and its effects become evident even after 30 $\Td$
(see also Bertin et al. 1994).

The result of the simulations are summarized in Fig. 1, where we plot
for all models the final value of the axis ratio, $\caf$, as a
function of $\xi$ (left panel) and $\ra/\rme$ (right panel) of initial
conditions (we recall here that $\rme$ is the spatial half--mass
radius).  A first result is that for one--component $\gm$-models (full
symbols) the $\xi$ critical value is in the range
$1.6\,\lsim\,\xis\,\lsim\,1.8$: for $\xi\,\lsim\,1.6$ all models were
found stable up to 100 $\Td$, while for $\xi\,\gsim\,1.8$ all models
present clear evolution on time--scales shorter than 30 $\Td$.  This
range for $\xis$ is compatible with the value 1.7, reported by FP84
and used in CL97, and with the results of Bertin et al. (1994), who
estimated $\xis\simeq 1.58$ for the family of ``$f_{\infty}$ models'';
Meza \& Zamorano (1997) found instead a higher threshold value for
stability $(\xis\simeq 2.3)$.  When expressed in terms of $\ra/\rme$
the critical value for stability is found in the range $0.6\,\lsim\,
(\ra/\rme)_{\rm s}\,\lsim\,0.8$.  Finally, the stability limits
expressed in terms of the (dimensionless) critical anisotropy radius
(the quantity of direct interest in this work) are given by
$\sas=(3.0,1.8,0.6)$ for $\gm=(0,1,2)$ models, respectively; for
comparison the critical anisotropy radius for consistency is
$\sac$=(0.5,0.2,0.0; Ciotti 1999).

For what concerns the internal structure of the end--products of
unstable initial conditions we found (in accordance with previous
results, see, e.g., Merritt \& Aguilar 1985, Stiavelli \& Sparke 1991)
that they are in general prolate systems, with axis ratios in the
range $0.3\,\lsim\,\caf\,\lsim\,1$, consistent with the ellipticities
of the observed galaxies. Only the most anisotropic models, near the
consistency limit $(\sa\simeq\sac$), form a triaxial bar. From Fig. 1
it is apparent that $\caf$ is strongly anti--correlated with $\xi$
(and so correlated with $\ra/\rme$), and this in a way essentially
independent of the value of $\gm$; a similar decrease of $\caf$ with
$\sa$ was also found in the numerical simulations of Meza \& Zamorano
(1997).

In order to better compare the end--products of unstable models with
real galaxies, we also fitted their projected mass density profiles
with the widely used Sersic (1968) $R^{1/m}$ law:
\begin{equation}
I(R)=I_0\,\exp\left[-b(m)\left(\frac{R}{\Re}\right)^{1/m}\right]
\end{equation}
where $b(m)\sim 2m-1/3+4/405m$ (Ciotti \& Bertin 1999).
\begin{figure}
\begin{center}
\parbox{1cm}{ \psfig{file=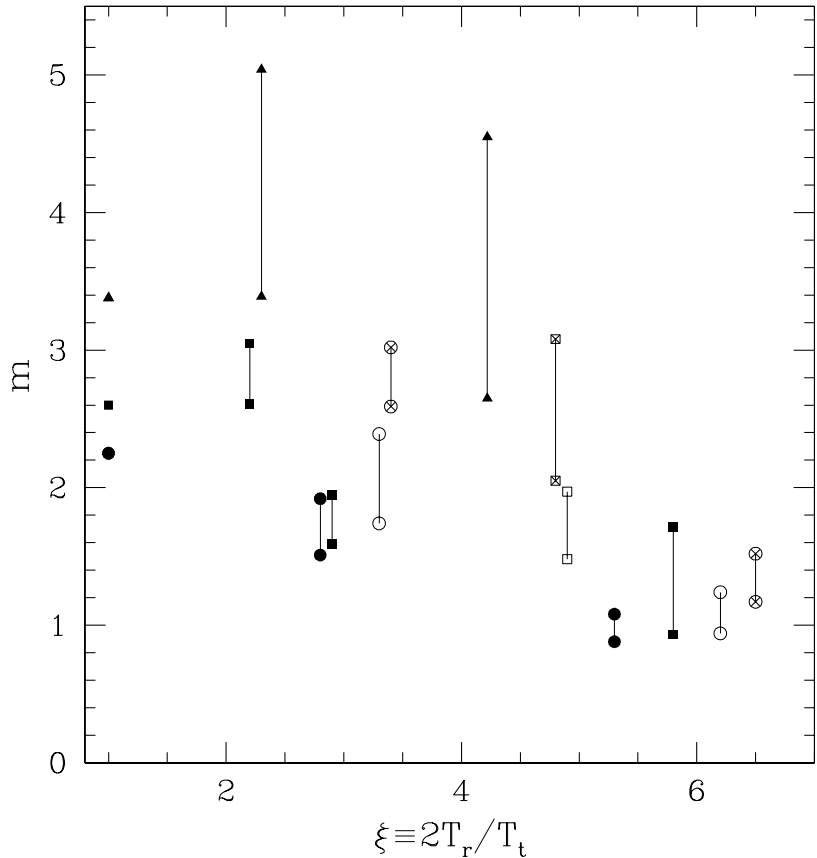,height=12cm}}
\caption{ Sersic best fit parameter $m$ vs. stability parameter $\xi$
for a subset of the computed models. Symbols are the same as in
Fig. 1. The three points at $\xi=1$ give the value of $m$ for the
isotropic (stable and spherically symmetric) parent galaxies, while
vertical bars show the maximum and minimum $m$ for each model due to
the relative orientation of the end--product and the line--of--sight.}
\end{center}
\end{figure}
In Fig. 2 we plot the Sersic best fit parameter $m$ as a function of
$\xi$ for a small, but representative, set of models. As for real
galaxies, we found a significant dependence of $m$ on the adopted
radial range over which the fit is performed, while the value of $m$
is not very sensitive to the specific fitting method adopted (see,
e.g., Bertin, Ciotti \& Del Principe 2001). For example, in the radial
range $0.1\,\lsim\, R/\cRe\,\lsim\, 4$ we found $1\,\lsim\, m\,\lsim\,
5$, with average residuals between the data and the fits
$\langle\Delta\mu\rangle\simeq 0.02\div 0.14\,{\rm mag}\,{\rm
arcsec^{-2}}$.  Clearly, the fitted quantities $m$ and $\cRe$ depend
on the relative orientation of the line--of--sight and of the
end--products of the simulations: with the vertical bars in Fig. 2 we
indicate the range of values spanned by $m$ when projecting the final
states along the short and long axis of their inertia ellipsoids.  As
an example of fit for a specific model, in Fig. 3 we show the data relative to the initial
conditions of an unstable $\gm=1$ model, and to the two projections
of its end--products.
\begin{figure}
\begin{center}
\parbox{1cm}{ \psfig{file=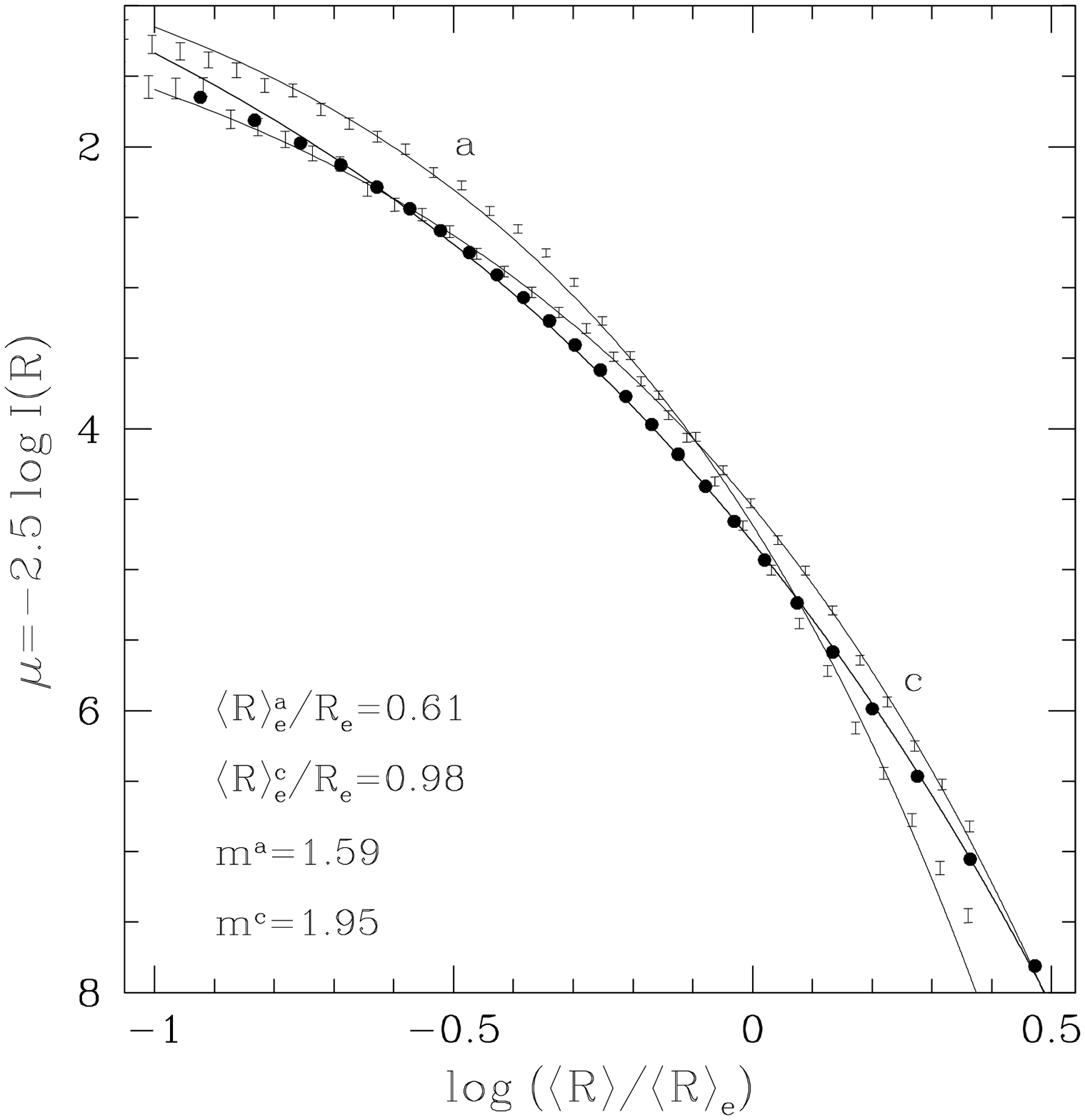,height=12cm}}
\caption{Circularized surface brightness profiles (vertical bars) of
the end--product of the unstable one--component $\gm=1$ model with
$\sa =0.7$. $\Re$ is the effective radius of the (spherically
symmetric) initial condition (solid dots), while
$\cRe\equiv\sqrt{\sae\sbe}$ is the circularized effective radius.
Solid lines represent the best fit Sersic models of the end--product
projections along its inertia ellipsoid minor ($c$) and major ($a$)
axis.}
\label{appenfig}
\end{center}
\end{figure}

Having determined for each family of galaxy models the critical
anisotropy radius, we can now proceed to check how distant from the FP
can stable models be placed, and where the end--products of unstable
initial conditions are found.  Of course, in this second case the
coordinates in $k$--space depend on the line--of--sight orientation
with respect to the density distribution of the end--products, and the
dependence is expected to be stronger for smaller values of $\caf$: to
any unstable initial condition corresponds, in the $k$--space, a {\it
set} of points.
\begin{figure}
\begin{center}
\parbox{1cm}{
 \psfig{file=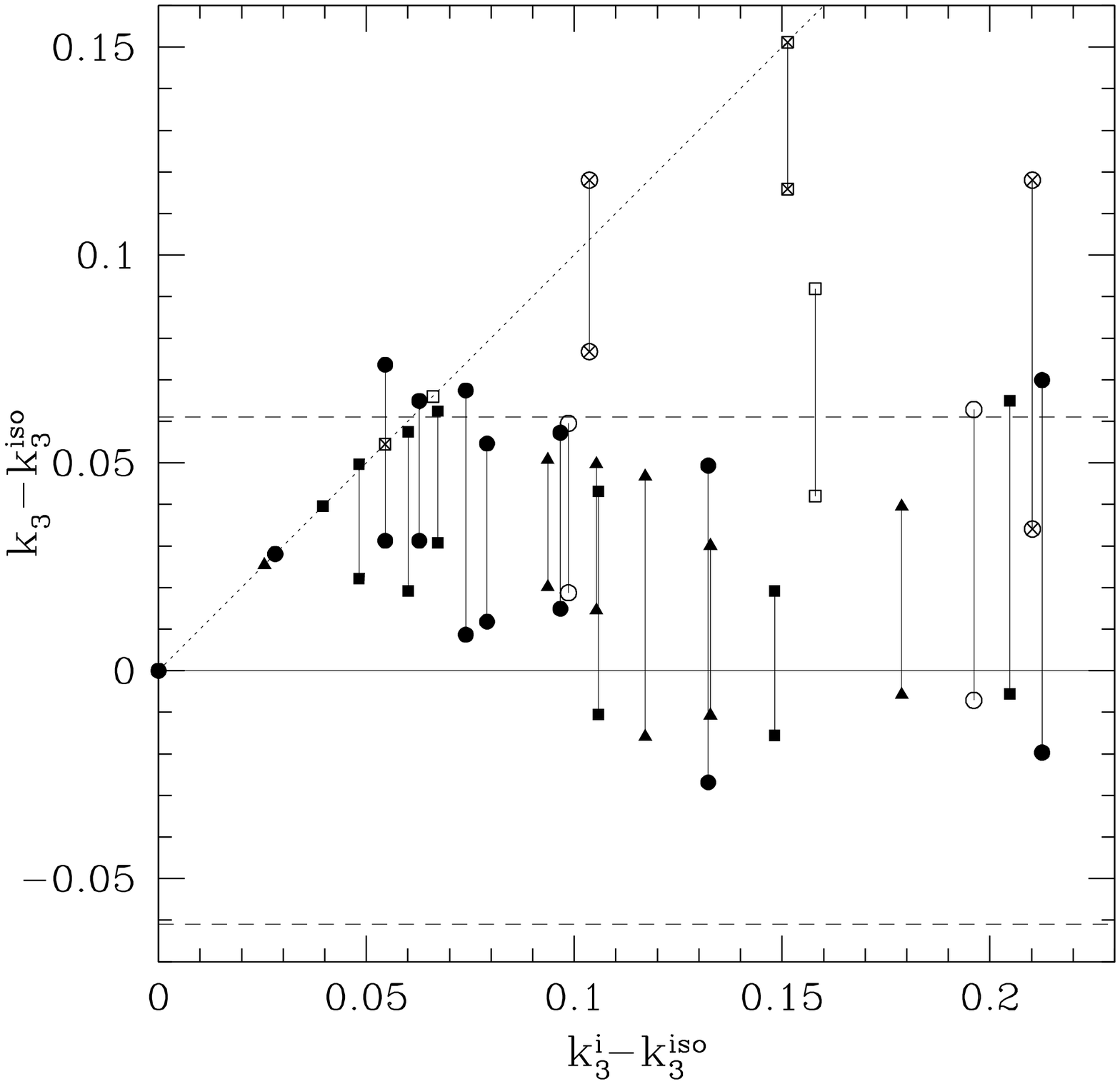,height=12cm}}
\caption{Final vs. initial $\kt$ for all galaxy models, measured with
respect to $\ktiso$, the $\kt$ coordinate of their isotropic parent
galaxy.  The horizontal dashed lines correspond to $\delta(\kt
)=\sigma(\kt)$, while the dotted line $\kt =\kti $ is the locus of the
initial conditions and of the stable models. Symbols are the same as
in Fig. 1. See the text for the explanation.}
\end{center}
\end{figure}
Due to the fact pointed out above that the properties of the models
here investigated do not depend on the specific position of the parent
galaxy on the FP, in Fig. 4 we plot the obtained results in a
coordinate system that reflects this property, and allows for an
immediate visualization of the most important consequences derived
from the simulations; in Fig. 5 the behavior of a representative set of
models is shown in the standard $(\ku ,\kd ,\kt)$ space. In the
horizontal axis of Fig. 4 we plot the displacement of the anisotropic
initial conditions, measured by their $\kti$, with respect to the
isotropic parent galaxy, placed at $\ktiso$: as a consequence, initial
conditions with larger $|\kti -\ktiso |$ are characterized by larger
amounts of radial orbital anisotropy.  As already discussed in the
Introduction, these initial conditions are placed in the $(\ku ,\kt)$
space along the lines given by equation (5). For example, in the upper
panel of Fig. 5 the initial conditions $\Au$, $\Bu$, and $\Cu$
correspond to the isotropic parent galaxies placed at points $A$, $B$,
and $C$ (by assuming $\ml =5$ and determining $\Lb$ from equation
[7]), while in the lower panel the same parent galaxies and initial
conditions are shown in the $(\ku ,\kd)$ space, connected by dotted
lines.

On the vertical axis of Fig. 4 we plot the quantity $\kt -\ktiso$
corresponding to the end--product of each explored initial
condition: if $\kt =\ktiso$ then the model has ``fallen back'' on the
FP. We also recall here that $\delta\kt\simeq 0.816|\kt-\ktiso|$: in
Fig. 4 the two horizontal dashed lines correspond to the FP thickness
$\sigma (\kt)/0.816\simeq 0.0613$, and so points inside this strip
represent models consistent with the observed thickness of the FP.
Finally, as an obvious consequence of the choice of the coordinate
axes in Fig. 4, note that all the {\it initial conditions} (as for
example models $\Au$, $\Bu$, and $\Cu$) are located at $t=0$ on the
dotted line (with slope equal to 1). This means that this line is also
the locus of {\it stable} models, while all the parent galaxies (as
for example models $A$, $B$, and $C$ of Fig. 5) lie at the point
$(0,0)$. A few stable anisotropic models may in fact be seen, as
single solid points, on this line at low anisotropy values.

By increasing the amount of radial anisotropy the initial conditions
move along the dotted line, and when they reach the critical value of the
anisotropy radius they become unstable and rearrange their density profile
and internal dynamics in a new, stable configuration. As we have seen,
these end--products are strongly asymmetric, and so their
representative points span a range of values as a function on the
line--of--sight orientation. In Fig. 4 the vertical lines show the
importance of this projection effect: note that, in general, the
length of these segments is considerably smaller than the FP
thickness.

The first result that can be obtained by inspection of Fig. 4 is the
fact that, for one--component $\gm$-models (solid symbols), radial
orbit instability becomes effective for initial conditions {\it
inside} the FP thickness, thus providing strong support to what found
by CL97 for $R^{1/m}$ models. On the contrary, by considering as
anisotropy limitation the very basic requirement of model consistency
only, it is apparent from Fig. 4 that physically acceptable initial
conditions could be placed at a distance from the FP substantially
larger than $\sigma (\kt)$, up to $\delta\kt\simeq 2\div3\sigma(\kt)$,
the well known problem motivating this work. Models $\Au$, $\Bu$, and
$\Cu$ in Fig. 5 are just three examples of such models.  The second
result is that all one--component, radially unstable models fall
back on the FP: in other words, {\it not only the FP thickness is
nicely related to stability, but the FP itself acts as an
``attractor'' for the end--products of radially unstable systems when
their parent galaxies lie on it}.

\begin{figure}
\begin{center}
\parbox{1cm}{
 \psfig{file=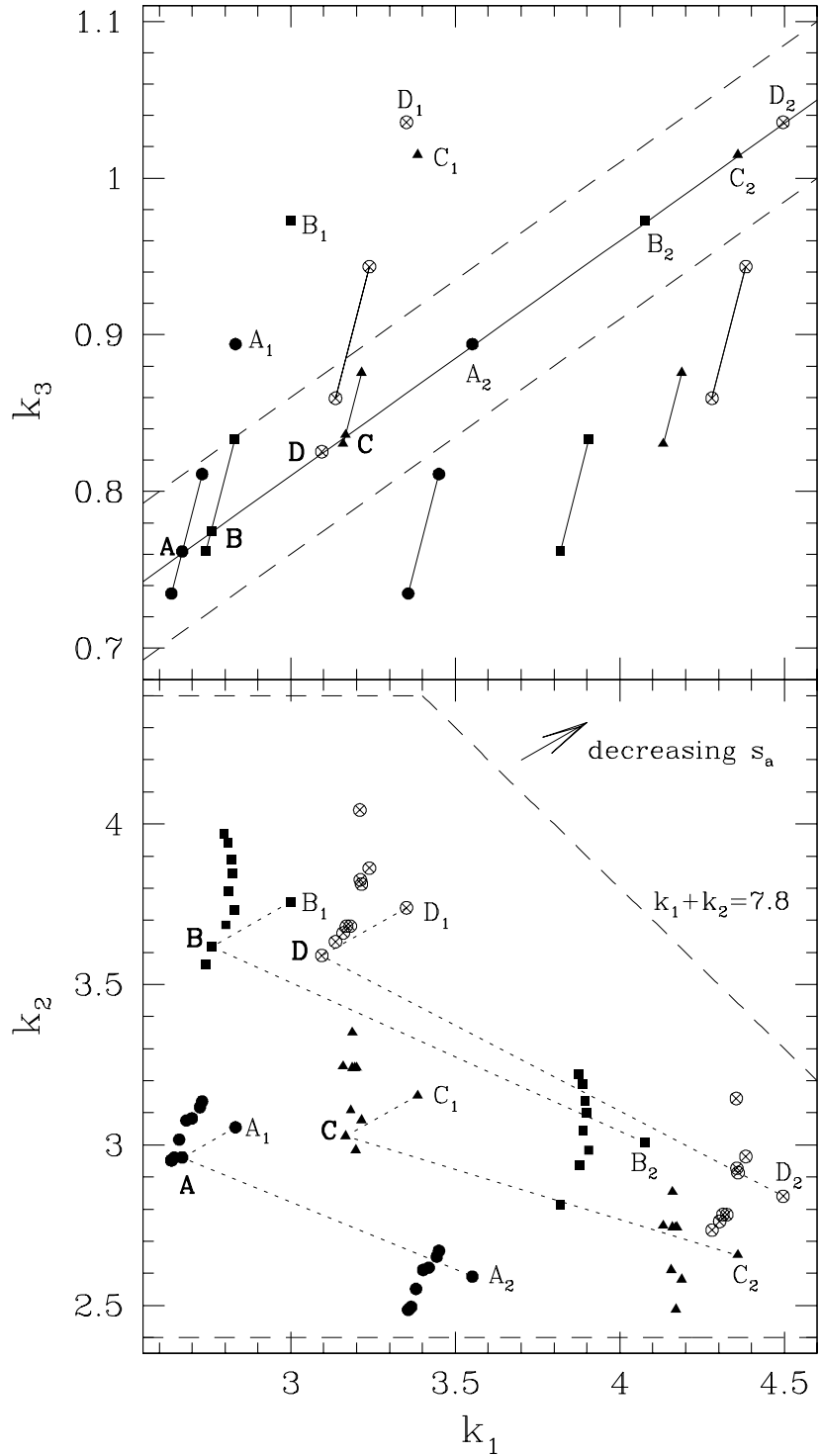,height=20.5cm}}
\caption{{\it Top}: $\kt$ vs. $\ku$ for a representative set of the
one and two--component models shown in Fig. 3 and Fig. 4: symbols are
the same. The solid line represents the FP relation (equation [4])
with its observed dispersion (dashed lines).  The stellar
mass--to--light ratio is fixed to $\ml=5$ and $\ml=2.7$, for one and
two--component models, respectively.  {\it Bottom}: $\kd$ vs. $\ku$
for the same models plotted in the upper panel. The dashed lines
define the region, as given by BBF, where real galaxies are found. The
arrow shows the direction followed by initial conditions with
increasing radial anisotropy.}
\end{center}
\end{figure}

The same results can be illustrated in a more direct way by using
Fig. 5, where we plot the positions of the end--products of
unstable models in the $k$ space.  As described above, to each final
configuration derived from unstable initial conditions corresponds a
set of points, depending on their relative orientation with respect to
the observer's line--of--sight. Due to the fact that the total
luminosity of each model is obviously conserved by projection, these
sets in the $(\ku ,\kt)$ space are actually {\it segments} of the
straight line given by equation (5), as shown by Fig. 5 where the
end--products of the initial conditions $\Au$, $\Bu$, and $\Cu$ can be
immediately recognized.  On the contrary, in the $(\ku,\kd)$ plane the
end--products are distributed, as a consequence of projection along
different angles, on two--dimensional regions. This is due to the fact
that no 1--1 relation similar to equation (5) exists between $\ku$ and
$\kd$, because also the value of $\cRe$ enters explicitly [$\kd=\log
(\ml\Lb^3/\Kv\cRe^6)/\sqrt{6}+const$]. For this reason for each
end--product we plot several positions obtained with random viewing
angles. Note however that the displacement in the $(\ku,\kd)$ space
occurs mainly along the $\kd$ coordinate, and this is due to the
steepening of the profile (Fig. 2), i.e., to the increase of
$\Ie$. {\it In any case, the models remain well inside the populated
zone of the $(\ku ,\kd)$ plane}.

\subsection{Two--component models}

We now present the results of the numerical simulations (8) of
two--component (1,1) galaxy models. In fact, although indications
exist that the {\it onset} of radial orbital anisotropy is not
strongly affected by the presence of a massive DM halo (see, e.g.,
Stiavelli \& Sparke 1991, Ciotti 1996), in principle its {\it
presence} could significantly modify the structural and dynamical
properties of the end--products of unstable initial conditions, and so
alter the findings obtained for one--component models described in
Section 3.1.

Unfortunately, due to the dimension of the parameter space, we are not
in the position to determine, as for one--component models, even a fiducial
threshold for stability, and so we limit our study to the behavior of
some representative models. In particular, we consider the following
cases: $\mu=1$ (``light'' halo), $\mu=3$ (``massive'' halo),
$\beta=0.5$ (``concentrated'' halo) and $\beta=2$ (``diffuse'' halo),
where for each of the four possible combinations we fix the
anisotropy radius to $\sa=0.3$ and $\sa=0.7$, two values corresponding
to strongly unstable one--component $\gm =1$ models.  In all the
simulations we use {\it live} DM halos, and in order to have equal
mass particles for ``stars'' and ``dark matter'' we adopt $N_{\rm
h}=98304$ and $N_*=32768$ for $\mu=3$, and $N_{\rm h}=N_*=65536$ for
$\mu=1$.

In analogy with the one--component case, the quantities $\Tr$ and
$\Tt$, entering the definition of $\xi=2\Tr /\Tt$, are now the radial
and the tangential component of the {\it total} (stellar and halo)
kinetic energy tensor, respectively.  This choice seems the
natural one in the case of a live DM halo, when from a
dynamical point of view the galaxy should be considered as a whole;
but certainly other choices (for example, by using in the $\xi$
definition the kinetic energies of the stellar component only) could
be equally well motivated. Clearly, the determination of the
observational properties of the end--products is based on the analysis
of their stellar component only.

As can be seen from Fig. 1, where the ellipticity of the stellar
component of (1,1) models is represented by empty symbols, the basic
trend of $\caf$ with $\xi$ is similar to that of one--component
models: the end--products are mostly prolate systems, with axis ratio
$\caf$ in the same range of that of the one--component models. In
general, however, when considering one and two--component initial
conditions with the same $\xi$, the final {\it stellar} distribution
remains more spherical in the cases with DM than in the one--component
cases.  Moreover, models with massive DM halos (empty symbols with
crosses in Figs. 1, 2, 4 and 5) remain more spherical than the
corresponding models with light halos and (approximately) the same
$\xi$.  This means that for the explored two--component models, at
variance with the one--component cases, the parameter $\xi$ {\it is
not} well correlated with the final axis ratio.  As can be seen from
Fig. 2, the best fit parameter $m$ of the projected stellar
distribution of the end--products of (1,1) unstable models remains
limited to values $m\,\lsim\,3$, the same range covered by
one--component $\gamma=1$ models. We also found that the final shape
of the DM halos remains nearly spherical, with
$0.88\,\lsim\,\caf\,\lsim\,1$.

The observational properties of the end--products of the
two--component models are illustrated in Figs. 4 and 5 where a
comparison with the one--component models can be easily made. In
particular, as in the one--component case, we found that the FP
thickness still nicely separates stable from unstable models,
independently of the amount and distribution of DM.  However, the
final position in the $k$ space of the end--products depends on the amount
of DM: as expected two--component models with a light halo (quite
independently of its concentration) are very similar to one--component
models, while models with massive DM halos are quite different. In
particular from Fig. 4 it is apparent that massive halos prevent the
models from falling back on the FP and this effect is stronger for more
concentrated halos. This is shown in Fig. 5 (upper panel) by the final
state of model $\Du$, an initial condition obtained from the parent
galaxy $D$, and characterized by a diffuse, massive DM halo. The
positions of this end--product in the $(\ku,\kd)$ plane are instead
remarkably similar to those of the one--component models, and the same
comments apply. Again, the scatter associated with projection effects
is smaller than the total thickness of the FP.

\section{Orbital anisotropy and the FP tilt}

As discussed in the Introduction, a question frequently addressed in
the literature is whether the so--called FP tilt can be due to some
kind of structural and/or dynamical ``non--homology'', i.e., to a
systematic variation from low to high luminosities of the structural
and/or dynamical properties of the galaxies. From an observational
point of view, this problem is still in general unsettled (see, e.g.,
Caon, Capaccioli \& D'Onofrio 1993, Graham \& Colless 1997, Gerhard et
al. 2001, Bertin et al. 2001 and references therein) and so here we
try to gain some hint on its solution by using the results of the
numerical simulations described in Section 3. In particular we
investigate whether relation (7) can be satisfied, in the whole
observed luminosity range ($0.2\, \lsim\, \Lb\, \lsim 40$ in the BBF
sample), by a systematic variation of $\Kv$ induced by an appropriate
underlying correlation $\sa=\sa(\Lb)$, while maintaining fixed the
models structure and $\ml$. In practice, the isotropic parent model of
each family is placed on the FP by selecting its $\Lb$ and $\ml$;
then, by increasing its anisotropy and luminosity, the family of
initial conditions is generated. Note that, at variance with the
exploration described in Section 3, in this case all the initial
conditions are placed by construction on the FP, as can be seen from
Fig. 5 (upper panel) where initial conditions $\Ad$, $\Bd$, $\Cd$ and
$\Dd$ are generated by the parent galaxies $A$, $B$, $C$ and $D$.

As in the case of Section 3, we found useful to represent the result
of the simulations in a coordinate system slightly different from the
usual $k$ space. In the abscissae axis of Fig. 6 we plot the quantity
$\kti -\ktiso$ that measures how much a given initial condition is
displaced on the FP from its parent isotropic model, while in the
ordinate axis we plot the quantity $\kt -\kti$ of the corresponding
end--product. As a consequence all the parent isotropic models are
placed at the origin, and the FP is represented by the solid line $\kt
=\kti$: this line is also the line of stable initial conditions, while
the two horizontal dashed lines represent the FP thickness. By using
an argument similar to that used in the case of Fig. 4, we now obtain
$\delta\kt\simeq 0.816|\kt-\kti|$ and the end--products of unstable
initial conditions are vertical segments due to projection effects.

\begin{figure}
\begin{center}
\parbox{1cm}{ \psfig{file=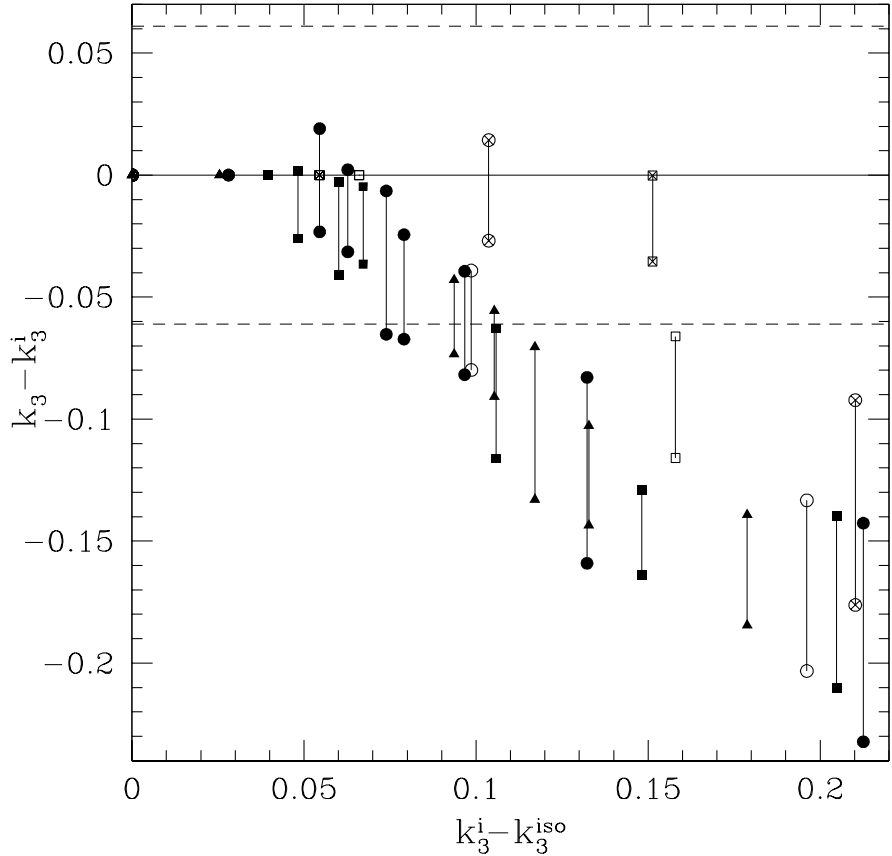,height=12cm}}
\caption{End--products of initial conditions placed on the FP at $\kti
-\ktiso$, where $\ktiso$ is the coordinate of their isotropic parent
galaxy. The dashed lines correspond to $\delta\kt =\sigma(\kt)$. Note
that the abscissae range is significantly smaller than the actual
range spanned by BBF galaxies, $\Delta\kt\simeq 0.3$. Symbols are the
same as in Fig. 4.}
\end{center}
\end{figure}

\subsection{One--component models}           

For each family of $\gm$-models we fixed a constant value of $\ml$
such that the isotropic model, with a suitable assigned $\Lb$, lies at
the faint end of the FP ($\ku \simeq 2.6$, $\kt\simeq 0.75$).  Then we
placed the anisotropic initial conditions on the FP by choosing $\Lb$,
according to equation (7), as a function of $\Kv=\Kv(\gm,\sa)$, in
order to reproduce the tilt of the FP.

The inspection of Fig. 6 reveals that it is not possible to reproduce
the FP tilt over the whole observed range ($\Delta\ku\simeq 2$ and
$\Delta\kt\simeq 0.3$ in the BBF sample) by using stable model only.
In fact, independently of the value of $\gm$, the variations of $\Kv$
for $\sa\gsim\sas$ correspond to $\Delta\kt\,\lsim\, 0.04$, much
smaller than the observed interval.  If we consider also unstable (but
consistent) systems, we can use a wider range of $\Kv (\gm,\sa)$
(cfr. Table 1). In this case it is possible to reproduce the FP tilt
over a much larger interval, $\Delta\kt \simeq 0.2$, which is however
still significantly smaller than the observed one: even if consistency
limitations only are taken into account, {\it the FP tilt cannot be
explained as an effect of a systematic increase of radial anisotropy
in one--component models, under the assumption of structural
homology} (see also CLR96 and CL97).

What happens to the end--products of unstable initial conditions is
shown in Fig. 6: in general they fall well outside the FP thickness,
and the departure is larger for larger distance from the parent
galaxy: they always fall out of $\sigma(\kt)$ for $\kti -\ktiso \simeq
0.1$: this implies that, limiting to end--products of unstable initial
conditions which remain {\it inside} the FP thickness, the maximum
amount of the FP tilt that can be explained by pure anisotropy 
is $\kti -\ktiso\, \lsim\, 0.1$. Of course, a {\it larger} part of the
FP tilt could be covered by considering parent galaxies with {\it
tangentially anisotropic} velocity dispersion tensor (as observed in
real, low luminosity ellipticals).

The results presented can also be seen, for a few representative
cases, in the upper panel of Fig. 5, where the position of the
end--products of models $\Ad$, $\Bd$, $\Cd$, $\Dd$ is represented by
the straight line segments. As in the case discussed in Section 3, the
end--products  remain well inside the region
populated by real galaxies in the $(\ku,\kd)$ plane.

Finally we qualitatively explore an interesting related problem (which
however is not the argument of our paper), i.e., whether the FP tilt
can be reproduced over the whole observed range in luminosity, for
fixed $\ml$, in case the requirement of perfect structural homology is
relaxed. In practice we determined $\ml$ so that the parent isotropic
$\gm=0$ model (to which corresponds the maximum value of $\Kv$) is
placed at the faint end of the FP ($\Lb=0.2$, $\ml=5$). With the same
value of $\ml$ the $\gm=1$ and $\gm=2$ parent galaxies are then placed
on the FP, and their positions are characterized by larger $\Lb$ (and
so $\ku$), due to the corresponding decrease of $\Kv$ (see first column
in Table 1). In this approach, the FP tilt can be reproduced over
almost all the observed range by using models compatible with {\it
consistency}. However, even in this {\it mixed} structural and
dynamical non--homology approach, if one limits to {\it stable} models
only, the maximum available range is reduced to $\Delta\kt \simeq
0.16$, approximately half of the required $\kt$ variation.
       
\subsection{Two--component models}           

In analogy with the one--component case, for each family of (1,1)
models, given $\mu$, $\beta$ (the same as for the models presented in
Section 3.2) and $\ml$, we placed the parent isotropic model at the
faint end of the FP and we derived $\Lb$ of the initial conditions
according to equation (7), as a function of
$\Kv=\Kv(\mu,\beta,\sa)$. As already discussed in Section 3.2, for
each family of (1,1) models we study only two values of the anisotropy
radius, $\sa=0.3$ and $\sa=0.7$.  We found that for each family these
initial conditions, under the only requirement of consistency
($\sa{\geq}\sac$), span a range $\Delta{k_3}\lsim0.2$, smaller than
the observed one (see Fig. 6). The main difference with respect to the
one--component case is that in the ($k_1$,$k_3$) space the
end--products of the unstable (1,1) models move less than the
one--component models with similar initial $\delta{k_3}$ and only the
most anisotropic systems fall outside the FP thickness. An example of
these systems is model $\Dd$, whose behavior in $(\ku,\kt)$ and
$(\ku,\kd)$ planes, clearly represented in upper and lower panel of
Fig. 5 respectively, is similar to that of one--component models
($\Ad$, $\Bd$, $\Cd$).  Again we found that the DM halo concentration
is an important quantity, strictly related to the displacement of the
end--product with respect to the initial conditions.

These findings suggest that {\it even in the two--component case under
the assumption of perfect structural homology and constant
mass-to-light ratio the FP tilt cannot be explained as a consequence
of a systematic increase of radial orbital anisotropy}.

\section{Discussion and conclusions}

With the aid of numerical simulations of one and two--component galaxy
models  we explored the constraints imposed by the
observed thickness and tilt of the FP on the amount and distribution
of radial orbital anisotropy in elliptical galaxies. The main results
are summarized below.

\begin{itemize}

\item Remarkably, all the explored models (both one and
two--component, and quite independently of the density profile) are
found to be unstable when their orbital radial anisotropy is high
enough to place them outside the observed FP thickness (under the
assumption that their isotropic parent models lie on the FP). On the contrary 
all stable models lie inside the FP thickness.

\item The end--products of one--component unstable models initially
placed outside the FP fall back inside the FP: in other words, the larger
is the initial displacement from the FP, the stronger is the
reassessment of the model structure and dynamics. The behavior of
two--component models is more varied, due to the fact that the
properties of their end--products are significantly affected by the amount of
mass and the distribution of the DM halo.  In particular, the
end--products of models with massive (either concentrated or diffuse)
DM halos remain outside the FP thickness, while models with light
halos behave essentially like one--component models.

\item Since the end--products of the unstable initial conditions are
not spherically symmetric, their positions on the $(\ku,\kt)$ plane
depend on their relative orientation with respect to the
line--of--sight direction.  However, the scatter due to projection
effects is in general smaller than the observed thickness of the FP,
both for one and two--component models.

\item We found that it is impossible to reproduce the whole FP tilt
with radially anisotropic but stable (one and two--component) models
under the assumption of constant $\ml$ and structural homology. In
other words, under these assumptions, luminous galaxies would be
radially unstable well before the bright end of the FP.

\item At variance with what happens to the end--products of unstable
models initially placed outside the FP thickness (but exactly for the
same reasons), the end--products of unstable models with initial
conditions on the FP, fall well outside the FP itself.

\end{itemize}


Our results lead to some speculation on the formation mechanism and
evolutionary scenarios of elliptical galaxies. First, if the (unknown)
formation mechanism produces galaxies with various degrees of internal
radial orbital anisotropy, of which the isotropic ones constitute the
``backbone'' of the observed FP, then the most anisotropic systems
would be radially unstable and would evolve into final states lying on
the FP. Then no ``ad hoc'' fine tuning would be required on the amount
of radial anisotropy of ellipticals at the moment of their formation.
In addition, our results concerning the FP tilt could give some
indications about the importance of dissipationless merging in the
history of the assembly of elliptical galaxies. In fact, if Es form by
hierarchical {\it dissipationless} merging, then the very existence of
the FP necessarily implies structural or dynamical non--homology of
the merging end--products. The possibility that we explored in this
work is that of a substantial dynamical non--homology as a function of
galaxy luminosity. Our simulations show that this is not a viable
possibility to reproduce the FP tilt: in this scenario the FP would be
destroyed by merging. However, it should be clear that we cannot rule
out the possibility that merging produces a combination of structural
and dynamical effects that conspire to maintain galaxies on the FP.
For this reason we are now exploring this problem, with the aid of one
and two--component galaxy merging simulations.

All our results on the FP thickness and tilt seem to point toward a
significant dynamical homology in real galaxies, and dynamical
homology in luminous Es has been recently determined by some authors
(Gerhard et al. 2001); we note also that an independent observational
support for dynamical homology is given by the very evidence of the
$M_{\rm BH}$-$\sg0$ relation, which relates a dynamical independent
quantity ($M_{\rm BH}$) with a quantity strongly dependent on
anisotropy ($\sg0$).  The fact that the scatter of the $M_{\rm
BH}$-$\sg0$ is very small means that elliptical galaxies are basically
dynamically homologous systems.

\section*{Acknowledgments}

L.C. would like to thank for useful discussion Ralf Bender, Giuseppe
Bertin, Roberto Saglia and Tjeerd van Albada.  C.N. and P.L. are
especially grateful to Lars Hernquist for having provided the
TREECODE; they also thank CINECA (Bologna) for assistance with the use
of T3E. We also thank Silvia Pellegrini for a careful reading of the
manuscript, and the referee, Massimo Stiavelli, for useful comments
that improved the paper. L.C. was supported by MURST CoFin2000 and by
ASI grant I/R/105/00.

\newpage

\appendix

\section[]{Intrinsic and observational properties of the numerical models}

In this Appendix we summarize the techniques used to derive the
intrinsic and projected properties of the end--products of the
simulations.

As discussed in Section 3.1, the intrinsic ellipticities associated
with the inertia ellipsoids of the end--products of the numerical
simulations were used as a measure of their departure from spherical
symmetry; thus for each numerical simulation we compute the inertia
tensor $I_{ij}=\sum_k x_i^{(k)}x_j^{(k)}$ associated with the density
distribution of interest (i.e., stars or halo).  Following Meza \&
Zamorano (1997), the sum is extended over all the particles inside
$r_{70}$, the radius of the sphere centered on the center of mass of
the galaxy and enclosing the 70\% of the total mass. The inertia
tensor is diagonalized and the ratios of the square root of its
eigenvalues are used to obtain a first estimate of the ellipticities
of the density distribution, according to the standard definition
$1-b_{70}/a_{70}$ and $1-c_{70}/a_{70}$. The orthogonal matrix
corresponding to the diagonalization is also obtained, and the density
distribution is rotated accordingly. This procedure is then applied
iteratively to the ellipsoid characterized by $a=r_{70}$ and with
intermediate and minor axes obtained from the ellipticities computed
at each stage, up to convergence at some prescribed accuracy level of
the ellipticities. Note that this procedure automatically select bound
particles only: in fact, in all our simulations the fraction of
``escapers'' never exceeds $0.02\%$ of the total number of particles.

In order to obtain the ``observational'' properties of our synthetic
galaxies, we calculate the following projected quantities of their
stellar component: the isophotal ellipticity $\epsilon$, the
circularized effective radius $\cRe$, the mean effective surface
brightness $\Ie$, and the central velocity dispersion $\sg0$.  The
line--of--sight direction is fixed by the arbitrary choice of
$\vartheta$ and $\varphi$, the two angles of spherical coordinates,
expressed in the reference frame where the inertia ellipsoid is
diagonal.  We apply to the system the rotation matrix
${\mathcal{R}}={\mathcal{R}}_2(\vartheta){\mathcal{R}}_3(\varphi)$,
where
\begin{equation}
{{\mathcal{R}}_3(\varphi)}=\left[
\begin{array}{ccc}
\mbox{cos}\varphi & \mbox{sin}\varphi & 0 \\
\mbox{-sin}\varphi & \mbox{cos}\varphi & 0 \\
0 & 0 & 1
\end{array}
\right],
\end{equation}
\begin{equation}
{{\mathcal{R}}_2(\vartheta)}=\left[
\begin{array}{ccc}
\mbox{cos}\vartheta & 0 &\mbox{-sin}\vartheta \\
0 & 1 & 0 \\
\mbox{sin}\vartheta & 0 &\mbox{cos}\vartheta  
\end{array}
\right].
\end{equation} 
In this coordinate system, the line--of--sight direction coincides
with the $z$ axis, while $(x,y)$ is the projection plane. The
isophotal axis ratio $b/a$ and the associated ellipticity
$\epsilon=1-b/a$ in the projection plane are determined by using a
two-dimensional version of the iterative scheme described above.

The semi-axes $\sae$ and $\sbe$, of the effective isophote are
determined under the assumption that the ellipticity of the projected
density is constant.  Finally, the circularized effective radius is
obtained:
\begin{equation}
\cRe=\sqrt{\sae\sbe}={\sae\sqrt{1-\epsilon}}.
\end{equation} 

From the knowledge of $\cRe$, the mean effective surface brightness
$\Ie{\equiv}\Lb/2\pi{\cRe}^2$ is derived modulo the free parameter
$\ml=M_*/\Lb$.

The ``central'' velocity dispersion $\sg0$ is computed, by restricting
to the central ellipse corresponding to a circularized radius
$\cRe/8$.  Thus we use
\begin{equation}
\sg0=\sqrt{\frac{1}{N}\sum_{i=1}^N\left(v_{z_i}-\bar{v}_z\right)^2},
\end{equation}
where $N$ is the number of particle in the projected ellipse of
semi-axes $\sae/8$ and $\sbe/8$, $v_{z_i}$ is the line--of--sight
velocity of the $i$-th particle, and $\bar{v}_z$ is the mean velocity
integrated along the line--of--sight. We find that discreteness
effects on the derived values of $\sg0$ range from $0.5\%$ (for
$\gamma=0$ models) to $0.9\%$ (for $\gamma=2$ models); moreover, the
uncertainties on $\cRe$ and $\Ie$ never exceed $0.7\%$ and $1.4\%$,
respectively. In other words, the numerical error bars on the model
measurements are significantly smaller than the observed scatter of
the FP itself, and so are not important in the context of the paper.

\bsp

\label{lastpage}

\end{document}